\documentclass[twoside,twocolumn,10pt]{article}
\usepackage{extsizes}
\usepackage[super,sort&compress,comma]{natbib} 
\usepackage[version=3]{mhchem}
\usepackage[left=1.5cm, right=1.5cm, top=1.785cm, bottom=2.0cm]{geometry}
\usepackage{balance}
\usepackage{comment}
\usepackage{mathptmx}
\usepackage{sectsty}
\usepackage{booktabs}
\usepackage{subcaption}
\usepackage{graphicx} 
\usepackage{orcidlink}

\usepackage{lastpage}
\usepackage[format=plain,justification=justified,singlelinecheck=false,font={stretch=1.125,small,sf},labelfont=bf,labelsep=space]{caption}
\usepackage{float}
\usepackage{fancyhdr}
\usepackage{fnpos}
\usepackage[english]{babel}
\addto{\captionsenglish}{%
  
}
\usepackage{hyperref}
\usepackage[nameinlink,capitalize]{cleveref}
\usepackage{array}
\usepackage{droidsans}
\usepackage{charter}
\usepackage[T1]{fontenc}
\usepackage[usenames,dvipsnames]{xcolor}
\usepackage{setspace}
\usepackage[compact]{titlesec}

\graphicspath{{./}{figures/}}
\usepackage{siunitx}

\usepackage{epstopdf}

\definecolor{cream}{RGB}{222,217,201}

\begin{document}

%%%PAGE SETUP%%%
\makeFNbottom
\makeatletter
\renewcommand\LARGE{\@setfontsize\LARGE{15pt}{17}}
\renewcommand\Large{\@setfontsize\Large{12pt}{14}}
\renewcommand\large{\@setfontsize\large{10pt}{12}}
\renewcommand\footnotesize{\@setfontsize\footnotesize{7pt}{10}}
\makeatother

\renewcommand{\thefootnote}{\fnsymbol{footnote}}
\renewcommand\footnoterule{\vspace*{1pt}% 
\hrule width 3.5in height 0.4pt \vspace*{5pt}} 
\setcounter{secnumdepth}{5}

\makeatletter 
\renewcommand\@biblabel[1]{#1}             
\renewcommand\@makefntext[1]% 
{\noindent\makebox[0pt][r]{\@thefnmark\,}#1}
\makeatother 
\renewcommand{\figurename}{\small{Fig.}~}
\sectionfont{\sffamily\Large}
\subsectionfont{\normalsize}
\subsubsectionfont{\bf}
\setstretch{1.125}
\setlength{\skip\footins}{0.8cm}
\setlength{\footnotesep}{0.25cm}
\setlength{\jot}{10pt}
\titlespacing*{\section}{0pt}{4pt}{4pt}
\titlespacing*{\subsection}{0pt}{15pt}{1pt}
%%%END OF PAGE SETUP%%%

%%%FOOTER SETUP FOR arXiv%%%
\pagestyle{fancy}
\fancyhead{}
\fancyfoot{}
\fancyfoot[C]{\footnotesize{\sffamily Page \thepage\ of \pageref{LastPage}}}
\renewcommand{\headrulewidth}{0pt} 
\renewcommand{\footrulewidth}{0pt}
\setlength{\arrayrulewidth}{1pt}
\setlength{\columnsep}{6.5mm}
\setlength\bibsep{1pt}
%%%END OF FOOTER%%%

%%%FIGURE SETUP (Without RSC Image Assets)%%%
\makeatletter 
\newlength{\figrulesep} 
\setlength{\figrulesep}{0.5\textfloatsep} 

\newcommand{\topfigrule}{\vspace*{-1pt}\hrule width \columnwidth height 0.4pt \vspace*{2pt}}
\newcommand{\botfigrule}{\vspace*{-2pt}\hrule width \columnwidth height 0.4pt \vspace*{2pt}}
\newcommand{\dblfigrule}{\vspace*{-1pt}\hrule width \textwidth height 0.4pt \vspace*{2pt}}
\makeatother
%%%END OF FIGURE SETUP%%%

%%%TITLE, AUTHORS AND ABSTRACT%%%
\twocolumn[
  \begin{@twocolumnfalse}
    \centering
    \sffamily
    
    \vspace*{0.5cm}
    {\LARGE \textbf{Curvature-Enhanced Polarization in Transition-Metal Dichalcogenide Janus Nanowrinkles}\par}
    
    \vspace{1.2em}
    {\large Stefan Velja\orcidlink{0009-0003-1268-6273},$^{a,*}$ Surender Kumar\orcidlink{0009-0000-3072-5633},$^{a}$ Domenico Corona\orcidlink{0000-0001-9053-9744},$^{a}$ and Caterina Cocchi\orcidlink{0000-0002-9243-9461}$^{a,b,\ddagger}$\par}
    
    \vspace{1.5em}
    \rule{\textwidth}{0.4pt}
    \vspace{0.8em}
\vspace{0.3cm}
    \begin{quote}
      \small
Strain-gradient engineering via out-of-plane wrinkling offers a powerful route to tune electronic and electromechanical properties in two-dimensional (2D) materials. Here, we systematically investigate the electronic and polarization response of wrinkled Janus \ce{MoSSe}/\ce{MoSeS} monolayers using density functional theory (DFT) calculations coupled with continuum elastica modeling. Exploring varying wrinkle sizes and compressive strain levels ($5\%$ -- $20\%$), we show that the global out-of-plane polarization follows a linear behavior when parameterized by the projected aspect ratio of the nanowrinkles. On this basis, we develop a physically grounded geometric model incorporating an effective 2D electromechanical coupling coefficient, which accurately predicts DFT polarizations without requiring higher-order nonlinear parameters. Atom- and orbital-resolved charge density analyses reveal the microscopic origin of this behavior: while the central Mo $4d$ projected weight remains approximately constant, local curvature drives continuous, chemically asymmetric charge transfer between the S $3p$ and Se $4p$ sublayer manifolds. Our findings establish clear geometric design rules for harnessing polarization in Janus monolayers for flexible nanoelectronics and optoelectronics.
    \end{quote}

    \vspace{0.5em}
    \rule{\textwidth}{0.4pt}
    \vspace{1.2em}
  \end{@twocolumnfalse}
]

\renewcommand*\rmdefault{bch}\normalfont\upshape
\rmfamily

\section*{}
\vspace{-1cm}
\footnotetext{\textit{$^{a}$~Institut für Festkörpertheorie und -Optik, Friedrich-Schiller-Universität Jena, 07743 Jena, Germany}}
\footnotetext{\textit{$^{b}$~Abbe Center of Photonics, Friedrich-Schiller-Universität Jena, 07745 Jena, Germany}}
\footnotetext{\textit{E-mail: $^{\ddag}$caterina.cocchi@uni-jena.de, $^{\ast}$stefan.velja@uni-jena.de}}

\section{Introduction}
\label{sec:introduction}
Engineering mechanical deformation in two-dimensional (2D) transition-metal dichalcogenides (TMDCs) has significantly advanced high-performance device applications, including single-photon emitters, photodetectors, and nanoelectronic systems~\cite{dai2019,naumis2017,rosenberger2019,maiti2020,akinwande2014}. Curved, out-of-plane nanoscale architectures such as wrinkles, folds, and bubbles~\cite{park2026,Katiyar2025,parto2021,kim2022,shabani2022} create spatially non-uniform strain patterns and large strain gradients~\cite{kumar2015,branny2016,Berry2019,cho2023}. Among these deformation modes, nanowrinkling has attracted particular attention because it generates continuous curvature profiles over large spatial domains. Strain gradients give rise to flexoelectricity, \textit{i.e.}, the generation of electric polarization in response to an applied strain gradient~\cite{Wenhao2018,xia2024}. Unlike piezoelectricity, flexoelectricity is universal to all dielectric materials regardless of crystallographic point-group symmetry~\cite{abdollahi2019}.

The local lattice distortions and polarization fields induced by flexoelectricity can substantially modify the electronic band structure, excitonic behavior, and charge-transport characteristics of 2D sheets~\cite{zheng2024,Iyengar2026}. Wrinkled TMDCs have demonstrated several strain-gradient-induced optoelectronic enhancements, including spatial energy funneling for high-efficiency solar harvesting, enhanced exciton dissociation, increased photocurrent density, and improved photovoltaic coefficients~\cite{kumar2021,codony2021}. Engineered wrinkle architectures have thus emerged as a robust platform for coupling mechanical deformations to electronic and optical functionality at the atomic limit.

Motivated by these advances, an important frontier is extending strain-gradient engineering to structurally asymmetric Janus TMDC monolayers~\cite{zhang2017,lu2017,Hu2024}. By breaking out-of-plane mirror symmetry at the atomic scale, Janus structures possess an intrinsic dipole moment~\cite{guan2018,ju2020,xia2018,torkashvand2026,kumar2026}. This has specific consequences for electronic and optoelectronic devices. Previous first-principles studies of Janus MXY/graphene heterostructures show that permanent Janus dipoles and interface-induced charge redistribution jointly control Schottky barriers and electrical contact types~\cite{chen2022role}. In monolayer Janus MoSSe, photocarrier lifetime under strain is primarily governed by the vertical dipole~\cite{fu2024vertical}, while in aligned Janus multilayers, intrinsic dipoles generate built-in electric fields and band offsets which are favourable for charge separation~\cite{riis2018efficient}. When combined with extrinsic curvature, Janus monolayers offer an unprecedented platform for manipulating electromechanical behavior. As the global aspect ratio dictates the overall spatial envelope, one might naively assume that this macroscopic structural parameter alone determines the polarization. This simplified picture, however, fails to capture how local bending modulates atomic-scale bond lengths and coordination angles, which represent the specific driving mechanisms for microscopic charge redistribution and atomic dipole modulation. Reconciling global morphology with local electromechanical coupling requires an accurate continuum profile description. Establishing whether local curvature modulates polarization through rigid atomic displacements or orbital-level charge redistribution remains an open question.

In this work, we investigate the electronic properties and curvature-enhanced polarization of wrinkled Janus \ce{MoSSe}/\ce{MoSeS} monolayers using density functional theory (DFT) calculations paired with continuum elastica modeling. We demonstrate that out-of-plane polarization follows an approximately linear trend with wrinkle aspect ratio and describe its geometry dependence using a projection model. By means of atom- and orbital-resolved charge density analysis, we uncover the microscopic origin of this response: while the Mo $4d$ projected weight varies comparatively little, local curvature drives an asymmetric charge redistribution between the $3p$- and $4p$-states of S and Se sublayers, respectively. Finally, we relate the polarization response to wrinkle geometry across the investigated sizes and compressions, providing a basis for strain-controlled polarization tuning in Janus-based nanoelectronics.

\section{Methods}
\label{sec:methods}
\subsection{Construction of the Systems and Ab Initio Computational Details}
\label{sec:computational}
\begin{figure}[t]
\centering
\includegraphics[width=\linewidth]{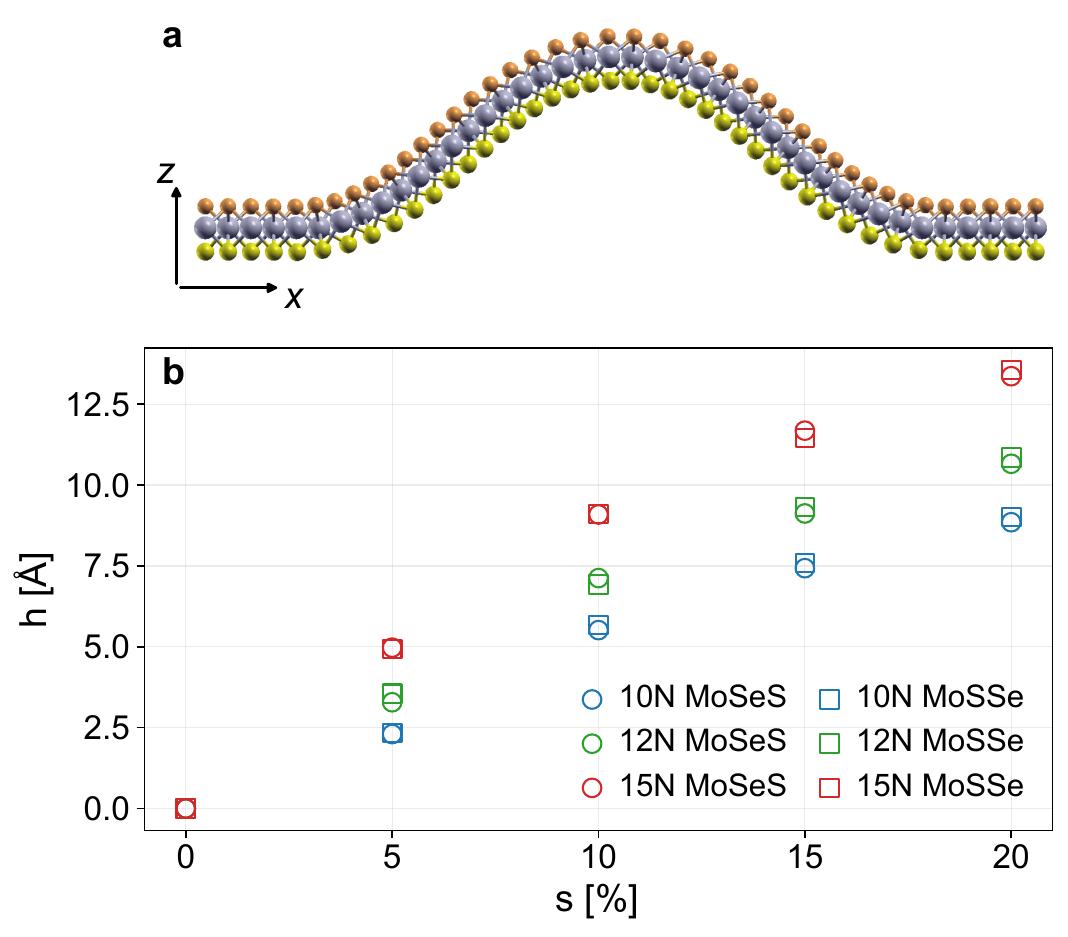}
\caption{(a) Ball-and-stick model of a relaxed $15N$ Janus MoSSe monolayer under $15\%$ global compressive strain. Mo atoms are depicted in gray, S atoms in yellow, and Se atoms in orange. (b) Maximum wrinkle height $h$ as a function of applied compressive strain $s$ across $10N$, $12N$, and $15N$ supercell sizes for both \ce{MoSeS} (open circles) and \ce{MoSSe} (open squares) orientations.}
\label{fig:struct}
\end{figure}

The nanowrinkles considered in this work were constructed from orthorhombic unit cells (OUCs) of monolayer Janus \ce{MoSSe}/\ce{MoSeS}, with optimized equilibrium lattice constants. Strained supercells were created by compressing $N \in \{10, 12, 15\}$ OUCs along the $x$-axis (Fig.~\ref{fig:struct}a) according to global strain values of $s \in \{5\%, 10\%, 15\%, 20\%\}$. The central strained region was padded with 3 fully relaxed, flat OUCs on each side to isolate the wrinkled zone from neighboring replicas~\cite{velja2024electronic}. The final wrinkle profiles were thus obtained by relaxing the free atomic coordinates under the imposed compression and the accompanying constraints. It should be noted that these results, as performed within the framework of DFT, characterize static electronic and polarization responses. They do not establish dynamic stability or finite-temperature persistence, which are outside the scope of this study.

The $c$ lattice constant of the supercell, along the $z$-axis, was set to 35~\AA{}, introducing a large vacuum along the out-of-plane direction to avoid spurious interactions between periodic images. This value is larger than what is typically found in the literature for the monolayers~\cite{mehdipour2022,sun2021armchair,shen2022thermal}, since we found out the dipole moment is particularly sensitive to the vacuum height. Furthermore, to prevent the propagation of the artificial electric field generated by the periodic repetition of polar Janus structures along the dipole direction $z$, we adopted the Bengtsson scheme~\cite{bengtsson1999dipole}, with the magnitude of the correction determined self-consistently from the evolving charge density. The maximum of the associated sawtooth potential was positioned so that the reversal region, with a width of $0.03c$, is entirely located in the vacuum region.

The DFT~\cite{Hohenberg1964,kohn1965} calculations reported in this work were performed using the plane-wave implementation provided by the \texttt{Quantum ESPRESSO} package~\cite{giannozzi2017}. Structural relaxations used scalar-relativistic Optimized Norm-Conserving Vanderbilt pseudopotentials~\cite{Hamann2013}, while all other calculations relied on fully relativistic pseudopotentials. Exchange-correlation interactions were treated within the generalized gradient approximation parameterized by Perdew, Burke, and Ernzerhof (PBE)~\cite{perdew1996generalized}. Kinetic energy cutoffs of 60~Ry and 240~Ry were applied to the electronic wavefunctions and the charge density, respectively. The Brillouin zone was sampled using a Monkhorst-Pack $k$-point mesh of $1 \times 8 \times 1$ centered at $\Gamma$. The convergence threshold for electronic self-consistency within each self-consistent cycle was set to $10^{-6}$~Ry. Geometry optimization was considered converged when the change in total energy between consecutive ionic relaxation steps fell below $10^{-4}$~Ry, with maximum residual force thresholds set to $10^{-5}$~Ry/bohr for the $10N$ supercells and to $10^{-4}$~Ry/bohr for $12N$ and $15N$.

\subsection{Global and Local Polarization Evaluation}
\label{sec:polarization}
We evaluate the macroscopic out-of-plane polarization $P_z$ of the nanowrinkles from the ionic core charges and the spatial electronic density $\rho(\mathbf{r})$:
\begin{equation}
P_z = \frac{d_z}{A_{\mathrm{arc}}}, \qquad d_z = \sum_a q_a z_a - \int \rho(\mathbf{r}) z\,\mathrm{d}V,
\label{eq:global-pz}
\end{equation}
where $q_a$ is the pseudopotential valence charge of ion $a$ at position $z_a$, and $A_{\mathrm{arc}} = L_{\mathrm{arc}} L_y$ is the mid-surface sheet area computed along the Mo-layer arc length $L_{\mathrm{arc}} = \int \sqrt{1 + (\mathrm{d}z/\mathrm{d}x)^2}\,\mathrm{d}x$ with transverse width $L_y$. To enforce charge neutrality, $\rho(\mathbf{r})$ is globally rescaled by a factor of $\sum_a q_a / \int \rho(\mathbf{r})\,\mathrm{d}V$, ensuring charge neutrality and origin invariance. Additionally, we reconstructed the real-space grid used for dipole integration from the full-precision cell vectors and integer grid dimensions, with matching full-precision ionic positions.

To interpret the microscopic polarization response, the electronic charge distribution is partitioned into atomic Voronoi basins $\{\Omega_a\}$~\cite{yu2011}. For each atom $a$ centered at position $\mathbf{R}_a$ with local surface normal $\hat{\mathbf{n}}_a$, the net atomic dipole moment relative to the local Mo-midsurface reference $\mathbf{R}_m$ is evaluated as:
\begin{equation}
d_a = \int_{\Omega_a} -\rho(\mathbf{r}) (\mathbf{r} - \mathbf{R}_a) \cdot \hat{\mathbf{n}}_a\,\mathrm{d}V + Q_a (\mathbf{R}_a - \mathbf{R}_m) \cdot \hat{\mathbf{n}}_a,
\label{eq:atom-dipole}
\end{equation}
where $Q_a = q_a - \int_{\Omega_a} \rho(\mathbf{r})\,\mathrm{d}V$ is the net Voronoi charge of the basin. The chalcogen dipoles ($d_{\mathrm{S}}$, $d_{\mathrm{Se}}$) are proportionally distributed to adjacent Mo formula units (FU) via distance-weighted allocation:
\begin{equation}
d_i^{\mathrm{FU}} = d_{\mathrm{Mo},i} + \sum_X w_{Xi} d_X, \qquad P_i^{\mathrm{FU}} = \frac{d_i^{\mathrm{FU}}}{A_i},
\label{eq:local-pz}
\end{equation}
where $A_i = A_{\mathrm{arc}} (V_i^{\mathrm{FU}} / \sum_j V_j^{\mathrm{FU}})$ scales the effective local area according to the Voronoi volume $V_i^{\mathrm{FU}}$ of formula unit $i$.

The Voronoi partitioning distributes the real-space electronic density among the atomic basins. To validate our tessellation choice, we compared our results with Bader analysis for all 24 wrinkles, and the two approaches are consistent in terms of the electron population in the chalcogens.

\subsection{Elastic Model for Wrinkle Reconstruction}
\label{sec:reconstruction}
To obtain a smooth continuum profile of the deformed mid-surface, the Mo-atom coordinates are parameterized using an elastica model. The local tangent angle $\theta(s)$ along the arc length $s$ is defined by a compact envelope profile:
\begin{equation}
\theta(s) = \theta_{\max} \frac{f_p(u)}{\max |f_p|}, \qquad f_p(u) = -u(1-u^2)^p \quad (|u| < 1),
\label{eq:elastica-theta}
\end{equation}
where $u = s/\ell$, with $\ell$ representing the characteristic length of the wrinkled area, and $\theta_{\max}$ is the peak tangent angle along the wrinkle flanks. The envelope exponent $p=1.629$ governs the envelope decay and is fixed across profiles to preserve geometric scaling.

The Cartesian midsurface profile is reconstructed via line integration:
\begin{equation}
x(s) = x_0 + \int_0^s \cos \theta(s')\,\mathrm{d}s', \qquad z(s) = z_0 + \int_0^s \sin \theta(s')\,\mathrm{d}s',
\label{eq:cartesian-reconstruction}
\end{equation}
where $x_0$ and $z_0$ are rigid translational offsets. This parameterization determines analytically the local curvature $\kappa(s) = \mathrm{d}\theta/\mathrm{d}s$ and the tangent vector $\mathbf{t}(s) = (\cos \theta, \sin \theta)$. The height and active projected length of the wrinkle are obtained from the reconstructed profile as:
\begin{equation}
h_{\mathrm{el}}=\max_{|s|\leq\ell}z(s)-\min_{|s|\leq\ell}z(s) \quad \text{and} \quad L_{\mathrm{el}}=\int_{-\ell}^{\ell}\cos\theta(s)\,\mathrm{d}s.
\label{eq:elastica-height-length}
\end{equation}

\subsection{Effective Electromechanical Model}
\label{sec:model_methods}
Upon out-of-plane corrugation, the S and Se layers of a Janus sheet lose structural equivalence, directly affecting the macroscopic polarization of the system~\cite{Sun2026}. To uniquely identify nanowrinkles, we adopt the convention where \ce{MoSSe} denotes the stacking with the S layer on top and Se on the bottom, while \ce{MoSeS} denotes the inverse stacking with Se on top and S on the bottom (Fig.~\ref{fig:struct}a).

To investigate both \ce{MoSSe} and \ce{MoSeS} nanowrinkles within a single, unified continuum framework, the direct-density polarization $P_z$ is sign-folded according to sublayer stacking:
\begin{equation}
P_z^{\mathrm{folded}} = \begin{cases} P_z, & \ce{MoSeS}, \\ -P_z, & \ce{MoSSe}. \end{cases}
\label{eq:folded-pz-def}
\end{equation}
The reconstructed total out-of-plane polarization $P_z^{\mathrm{rec,folded}}$ is represented by a linear geometric projection model:
\begin{equation}
P_z^{\mathrm{rec,folded}} = a + b X,
\label{eq:phys-flexo-model}
\end{equation}
\begin{equation}
X = \frac{N_{\mathrm{FU}}}{A_{\mathrm{arc}}} \left\langle \cos(\theta) \right\rangle,
\label{eq:projection-descriptor}
\end{equation}
where $\langle \dots \rangle$ denote spatial averages evaluated across the deformed wrinkle profile.

The factor $\cos(\theta)$ is the $z$-axis component of the local sheet normal vector, and describes the geometric projection of a normal-directed dipole. In the term $bX$, the factor $b N_{\mathrm{FU}} \langle \cos(\theta) \rangle$ represents an effective projected dipole moment. Division by $A_{\mathrm{arc}}$ converts this dipole contribution into polarization. The mean uses the midpoint normals of the Mo layer, weighted by their spline arc lengths, including the padding. The parameters $a$ and $b$ are obtained by least squares and have units of $e/\mathrm{\mathring{A}}$ and $e\mathrm{\mathring{A}}$, respectively.

\section{Results}
\label{sec:results}
In investigating the electronic properties and curvature-induced polarization response of wrinkled Janus \ce{MoSSe}/\ce{MoSeS} monolayers, we first establish the structural scaling and elastica reconstruction under compressive strain. Next, we analyze the strain dependence of the electronic band structure and orbital hybridization. We then present the global out-of-plane polarization response and validate our continuum geometry prediction model. Finally, we resolve the microscopic origin of the polarization response through atom- and orbital-resolved charge density analysis.

\subsection{Wrinkle Morphology and Mechanical Response}
\label{sec:geometry}
Under global compressive strain $s \in \{5\%, 10\%, 15\%, 20\%\}$, the Janus monolayers undergo out-of-plane buckling to form periodic, smooth nanowrinkles. Across all investigated supercells ($10N$, $12N$, and $15N$), the maximum wrinkle height $h$ increases sublinearly with applied strain $s$ (Fig.~\ref{fig:struct}b). At fixed strain, longer supercells attain larger peak amplitudes. Notably, the height profiles for \ce{MoSeS} and \ce{MoSSe} overlap almost identically, confirming that the mechanical buckling is symmetric with respect to chalcogen layer inversion.

The atomic coordinates of the relaxed Mo-layer are parameterized using our tangent-angle continuum elastica model~\cite{levien2008elastica}, which captures the relaxed mid-surface profile with extreme fidelity across the entire dataset of 24 relaxed configurations ($R^2 > 0.999$, coordinate root mean squared error (RMSE) $< 0.06$~\AA{}). This structural agreement confirms that the continuous geometric descriptors $\kappa(s)$ and $\theta(s)$ faithfully represent the atomistic mid-surface without introducing shape discretization noise into the spatial averages.

\subsection{Electronic Band Structure and Orbital Rehybridization}
\label{sec:electronic}
\begin{figure}[t]
\centering
\includegraphics[width=\linewidth]{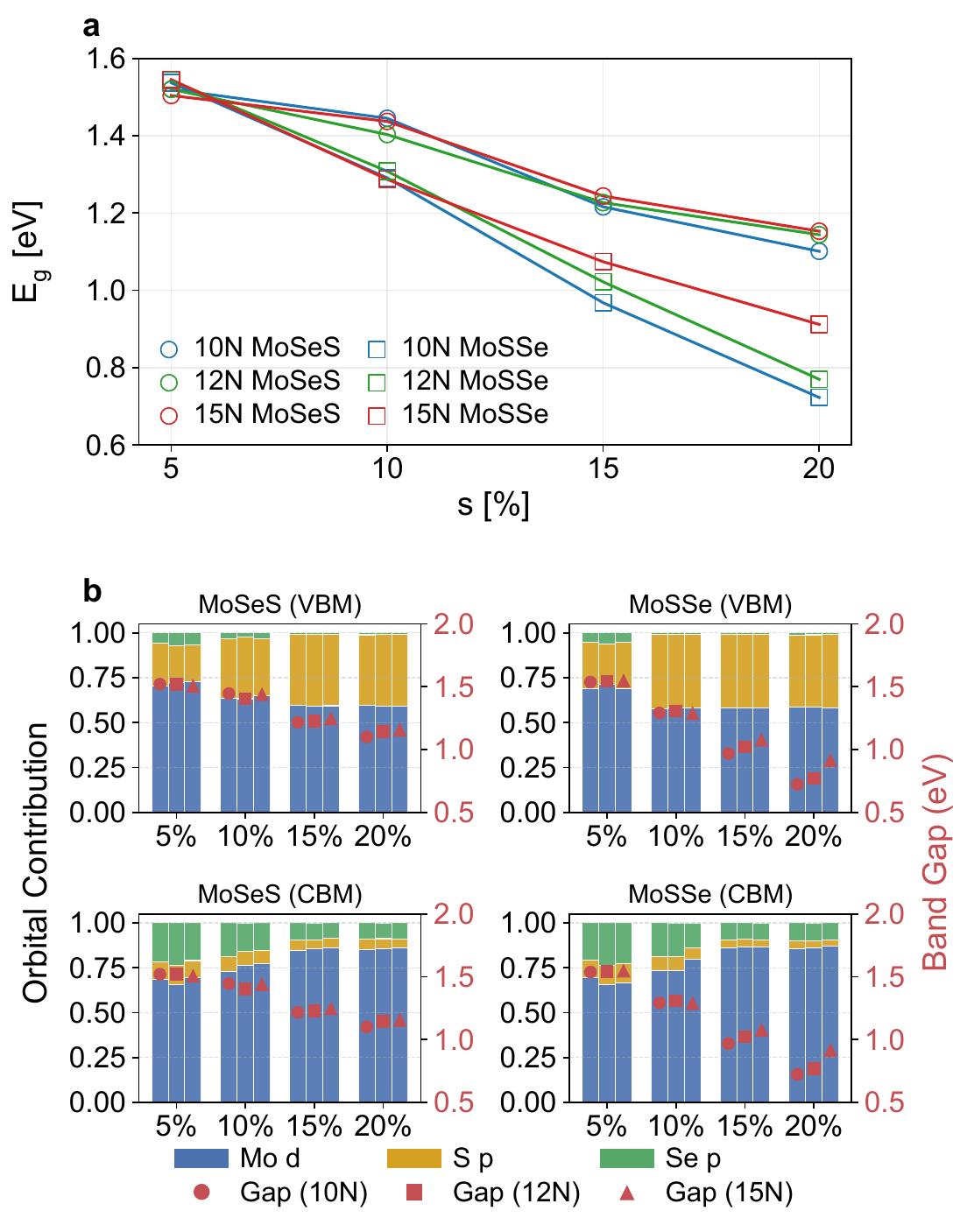}
\caption{(a) Fundamental band gap $E_g$ as a function of compressive strain $s$ for \ce{MoSeS} and \ce{MoSSe} nanowrinkles in their $10N$, $12N$, and $15N$ supercells. (b) Orbital contributions (Mo $d$, S $p$, Se $p$) to the valence band maximum (VBM) and conduction band minimum (CBM) as a function of strain (colored bars) alongside the band-gap values (red symbols).}
\label{fig:pdos-band-gaps}
\end{figure}

Structural wrinkling significantly modulates the electronic structure of TMDCs~\cite{qi+13apl,yu2016bending,Lee2023,Zhang2024,velja2024electronic}, which is further enhanced in Janus monolayers due to their broken intrinsic mirror symmetry~\cite{Yagmurcukardes2020,Zhang2022}. As shown in Fig.~\ref{fig:pdos-band-gaps}a, the fundamental gap decreases monotonically with increasing compressive strain across all supercell sizes maintaining, however, a markedly orientation-dependent magnitude.

Under a moderate strain of $s = 5\%$, all configurations exhibit comparable band gaps on the order of 1.5~eV. While the PBE calculations performed here notoriously underestimate absolute band gaps~\cite{perdew2001jacob}, the relative trends and energetic shifts across the examined structural configurations remain quantitatively robust. As the compressive strain increases to 10\%, a clear bifurcation emerges between the two datasets. The band gap of \ce{MoSeS} nanowrinkles remains at or above 1.4~eV, whereas in their \ce{MoSSe} counterparts, it drops sharply to about 1.3~eV. The trend intensifies at $s = 15\%$, where the MoSeS structures maintain gaps around 1.2~eV, compared to 1.1--0.9~eV for MoSSe. Under the largest deformation of 20\%, the band gap of \ce{MoSeS} contracts to $\sim$ 1.1~eV, while the \ce{MoSSe} gap reduces further to 0.7--1.0~eV.

To elucidate the microscopic origin of this orientation-asymmetric band-gap closure, we project the wavefunctions at the valence band maximum (VBM) and conduction band minimum (CBM) onto atomic orbital bases (Mo $d$, S $p$, and Se $p$) (Fig.~\ref{fig:pdos-band-gaps}b).

Consistent with the flat Janus monolayer reference~\cite{mehdipour2022}, the CBM is predominantly composed of Mo $d$ orbitals across all strain regimes. S $p$-states contribute a minor fraction ($\le 10\%$), whereas the weight of Se $p$-orbitals steadily diminishes under increasing compression, to the advantage of the Mo $d$-states.

Conversely, the VBM exhibits a distinct orbital reconfiguration under deformation. Se $p$ states contribute only at low strain ($s = 5\%$) before vanishing at larger flexure, whereas the weight of the S $p$ orbitals grows with strain, rising from $\sim 20\%$ in mildly deformed profiles to nearly $40\%$ at $s = 20\%$. Remarkably, this hybridization is governed almost entirely by local curvature and strain, remaining largely insensitive to global supercell periodicity.

The orientation asymmetry in gap narrowing stems from the differential strain environment imposed on the top and bottom chalcogen sublayers. In \ce{MoSSe} wrinkles, the outer convex crest places the smaller, more electronegative S atoms under local tensile strain while compressing the larger Se atoms at the inner concave trough. This tensile deformation at the apex elevates the energetic position of the S $p$--Mo $d$ hybrid states at the VBM, accelerating band gap closure compared to the inverted \ce{MoSeS} geometry.

\begin{figure}[t]
\centering
\includegraphics[width=\linewidth]{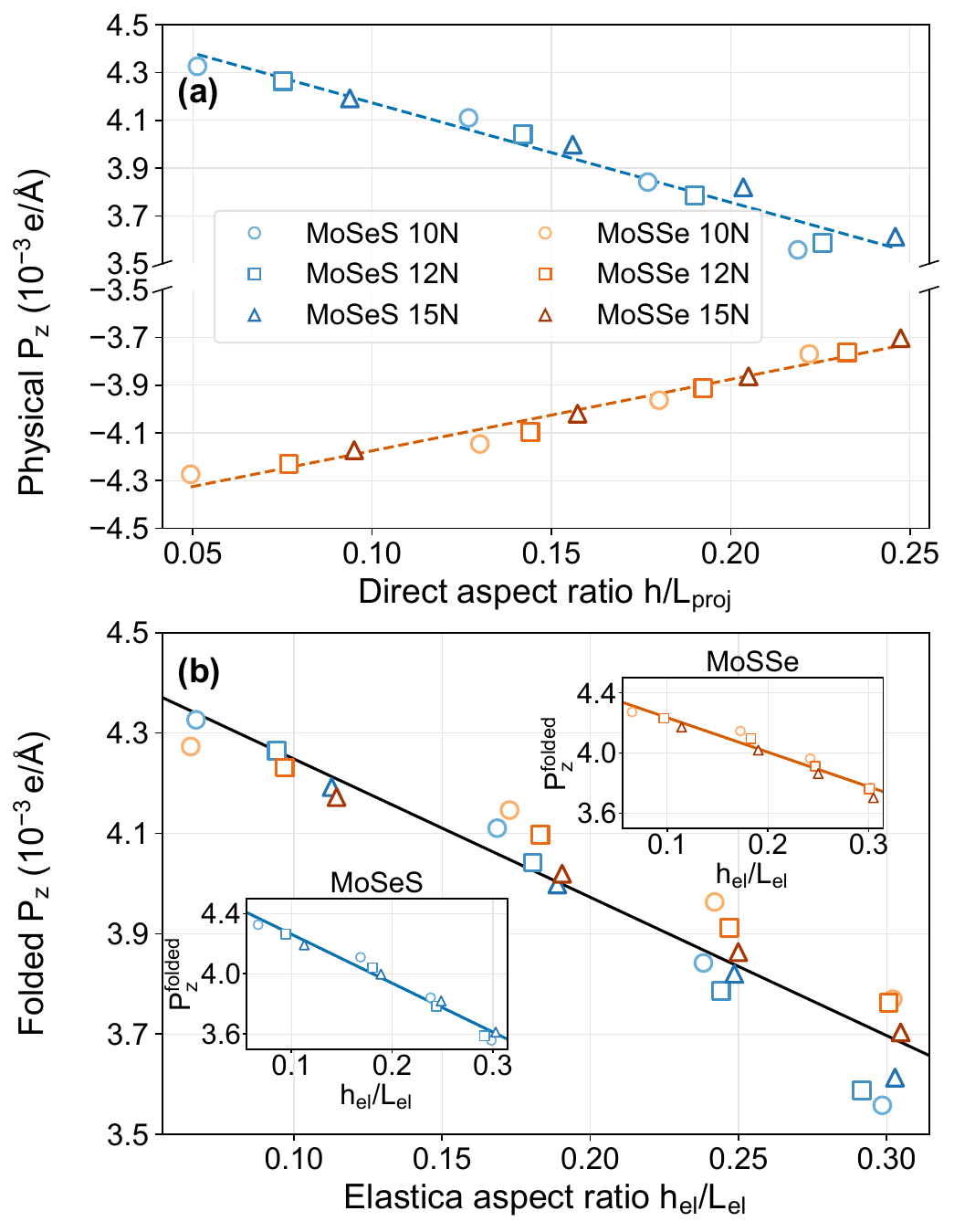}
\caption{(a) Macroscopic polarization $P_z$ versus the aspect ratio $h/L_{\mathrm{proj}}$ of the nanowrinkles. (b) Direct-density sign-folded polarization $P_z^{\mathrm{folded}}$ versus the elastica aspect ratio $h_{\mathrm{el}}/L_{\mathrm{el}}$. The solid black line denotes the pooled linear fit across all $10N$, $12N$, and $15N$ regular structures, while the insets showcase separate fits for \ce{MoSSe} and \ce{MoSeS}.}
\label{fig:global-pz}
\end{figure}

\subsection{Global Polarization and Continuum Model Validation}
\label{sec:global_polarization}
The macroscopic polarization $P_z$ assesses the out-of-plane electromechanical response of the Janus nanowrinkles. Prior to geometric scaling (Fig.~\ref{fig:global-pz}a), $P_z$ splits into two symmetric branches (\ce{MoSeS} \textit{vs.} \ce{MoSSe}) and displays a clear dependence on both aspect ratio and supercell size. However, applying our sign-folding convention (Eq.~\ref{eq:folded-pz-def}) and plotting the whole dataset against the aspect ratio of the elastica model, $h_\mathrm{el}/L_{\mathrm{el}}$, the 24 data points follow an approximately linear trend (Fig.~\ref{fig:global-pz}b). This collapse confirms that the net out-of-plane polarization is governed predominantly by the projected wrinkle geometry rather than supercell size effects. However, it is yet to be established to what extent can this be generalized to structures, whose compositions, size, or type of deformation are not within the scope of this study.

To bridge the continuum morphology with the DFT results, we parameterize $P_z^{\mathrm{rec,folded}}$ using the linear electromechanical model introduced in Eq.~\ref{eq:phys-flexo-model}. The universal linear collapse seen in Fig.~\ref{fig:global-pz}b directly reflects the continuum profile, where the averaged curvature scales as $\langle |\kappa| \rangle \propto h_{\mathrm{el}}/L_{\mathrm{el}}^2$. Parameterizing Eq.~\ref{eq:phys-flexo-model} across all 24 configurations confirms that the electromechanical response is linear with respect to $\langle |\kappa| \rangle$, achieving a coefficient of determination $R^2 = 0.931$ (Table~\ref{tab:flexo_fit}).

\begin{table}[h]
\centering
\caption{Fitted parameters of the linear electromechanical model (Eq.~\ref{eq:phys-flexo-model}), and the coefficient of determination ($R^2$) evaluating the quality of the linear fit.}
\label{tab:flexo_fit}
\small
\begin{tabular}{lccc}
\hline\hline
Dataset & $a$ ($10^{-4}\,e/\mathrm{\mathring{A}}$) & $b$ (Debye) & $R^2$ \\
\hline
\ce{MoSeS} & -3.3844 & 0.2027 & 0.985 \\
\ce{MoSSe} & +9.6097 & 0.1444 & 0.986 \\
All (Pooled) & +3.3936 & 0.1723 & 0.931 \\
\hline\hline
\end{tabular}
\end{table}

\begin{figure}[t]
\centering
\includegraphics[width=\linewidth]{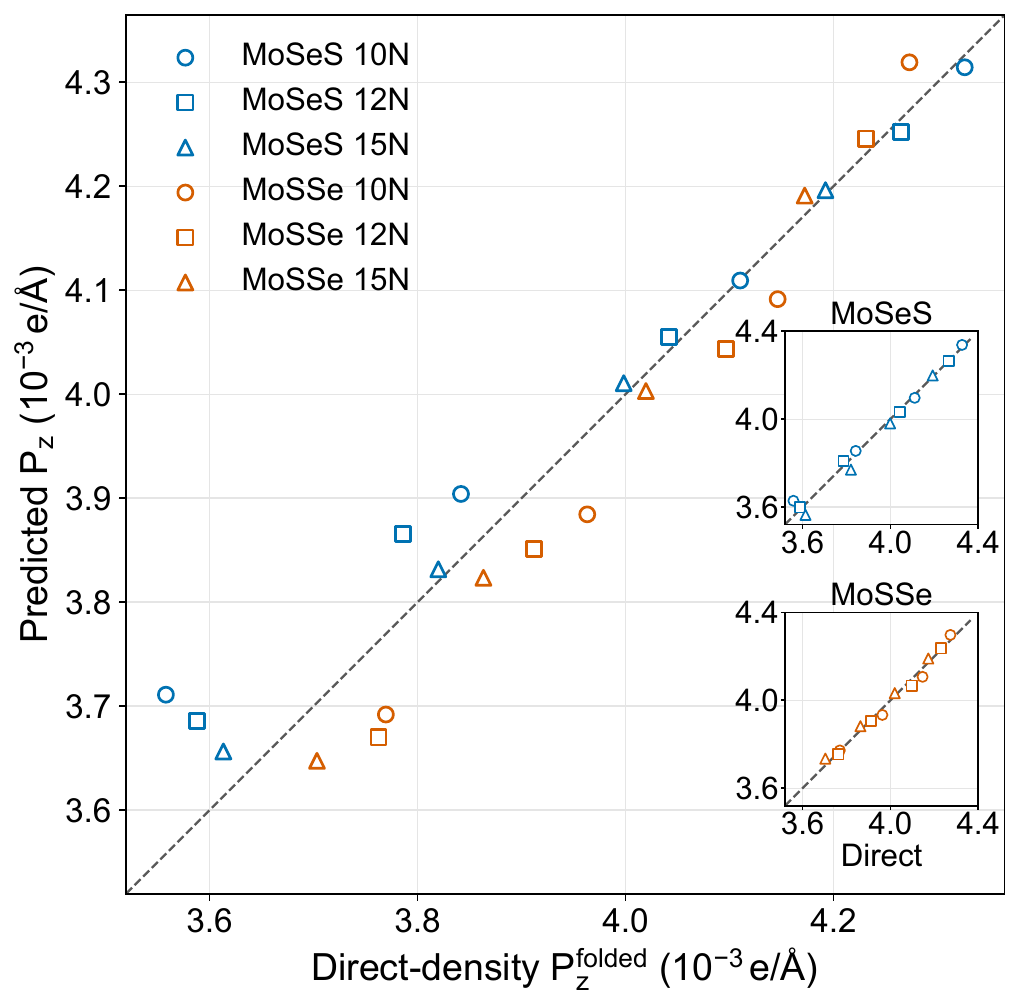}
\caption{Comparison of predicted polarization $P_z^{\mathrm{rec,folded}}$ from the effective electromechanical model against DFT direct-density calculations $P_z^{\mathrm{folded}}$. The insets showcase separate predictions for \ce{MoSSe} and \ce{MoSeS}.}
\label{fig:prediction}
\end{figure}

Independent fits to the two Janus orientations give slopes of $b=0.2027$~Debye for \ce{MoSeS} and $b=0.1444$~Debye for \ce{MoSSe}, with $R^2=0.9854$ and $0.9856$, respectively. Each orientation-specific fit contains 12 configurations, whereas the pooled fit uses a common parameter pair for all 24 configurations. As shown in Fig.~\ref{fig:prediction}, the pooled reconstructed polarizations agree with the DFT values with an RMSE of $5.91\times10^{-5}~e/\mathring{A}$.

\subsection{Microscopic Polarization and Inter-Chalcogen Charge Transfer}
\label{sec:microscopic_mechanism}
\begin{figure}[t]
\centering
\includegraphics[width=\linewidth]{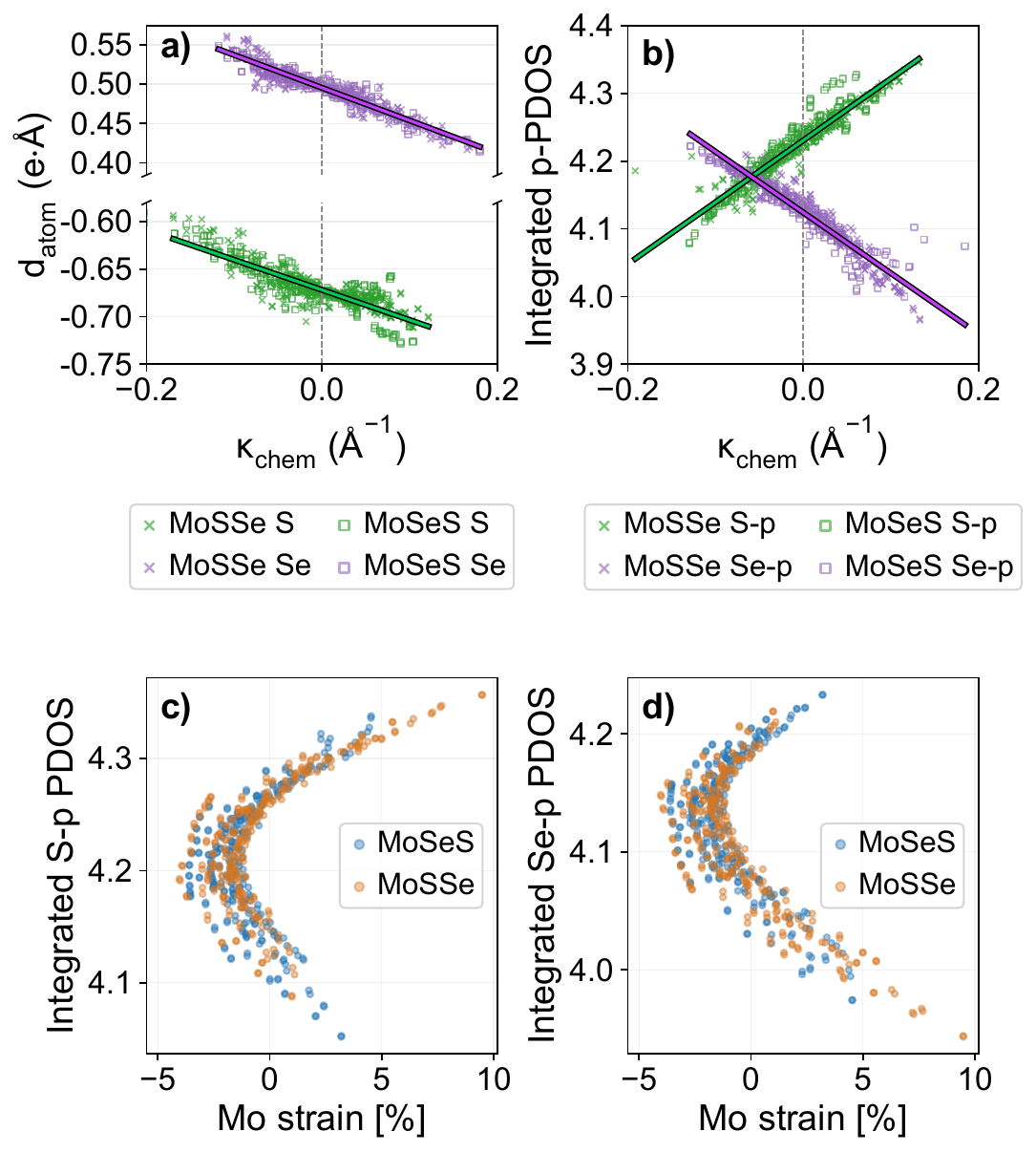}
\caption{(a) Local atomic dipole moment $d_{\mathrm{atom}}$ ($e\cdot\mathrm{\mathring{A}}$) for S and Se atoms as a function of local curvature $\kappa$. (b) Integrated $p$-orbital electron occupations ($N_e$) in the S $3p$ and Se $4p$ manifolds versus local curvature $\kappa$. (c,d) Integrated S-$p$ and Se-$p$ PDOS, respectively, plotted against the longitudinal strain of the surrounding Mo sublattice. The strain is obtained as the average over the three nearest Mo neighbours and referenced to the inter-metal distance in pristine MoSSe.}
\label{fig:local-chalcogen}
\end{figure}

To uncover the microscopic origin of the electromechanical coupling, we evaluate the atom-resolved local dipoles ($d_{\mathrm{atom}}$) and orbital occupations as a function of the chemically signed curvature $\kappa$. We recall that, according to our convention, negative curvature ($\kappa < 0$) denotes convex bending relative to the top chalcogen layer (outer crest), placing it under local tensile strain, while positive curvature ($\kappa > 0$) corresponds to the inner concave trough under compression. As shown in Fig.~\ref{fig:local-chalcogen}a, the atomic dipoles associated with the sulfur and selenium sublayers occupy distinct baseline regimes due to the intrinsic Janus charge asymmetry: $d_{\mathrm{atom}}$ is consistently positive for Se and negative for S. Upon bending, both atomic dipole moments exhibit parallel, linear decreases as a function of $\kappa$, confirming that local mechanical corrugation modulates charge within the individual atomic Voronoi basins continuously and predictably across both \ce{MoSSe} and \ce{MoSeS}.

Out-of-plane bending induces a continuous, reversible inter-chalcogen charge redistribution within the $p$-orbital manifolds (Fig.~\ref{fig:local-chalcogen}b). As the local curvature $\kappa$ increases from convex ($\kappa < 0$) to concave ($\kappa > 0$), there is a net charge transfer from the Se $4p$- to the S $3p$-orbital manifold. This redistribution is strictly linear, characterized by an electron accumulation rate of $\mathrm{d}N_{e,\mathrm{S}3p}/\mathrm{d}\kappa = +0.913~e\cdot\mathrm{\mathring{A}}$ in the S $3p$ states and a symmetric depletion rate of $\mathrm{d}N_{e,\mathrm{Se}4p}/\mathrm{d}\kappa = -0.898~e\cdot\mathrm{\mathring{A}}$ in the Se $4p$ states. Importantly, the crossing between the S $3p$ and Se $4p$ orbital occupation curves at $\kappa_{\mathrm{chem}} \approx -0.05~\mathrm{\mathring{A}}^{-1}$ marks the point where mechanical deformation compensates the intrinsic dipole moment of Janus monolayers, restoring local orbital charge symmetry between the two chalcogen sublayers.

Since the central Mo $4d$ electron occupation remains essentially invariant under bending, the transition-metal sublayer provides a stable, rigid contribution to the curvature-dependent response. Meanwhile, the asymmetric chalcogen sublayers provide a dynamic, curvature-driven tuning mechanism governed by localized $p$-orbital rehybridization. This microscopic polarization mechanism complements macroscopic scaling and linearity established in Section~\ref{sec:global_polarization}.

We additionally observe the connection between the integrated PDOS of the chalcogen p-orbitals and the longitudinal strain among the surrounding Mo atoms. When expressed as strain versus integrated PDOS, this relationship is approximately parabolic (Fig.~\ref{fig:local-chalcogen}c,d).

\section{Discussion}
\label{sec:discussion}
The coupling between out-of-plane curvature and global polarization demonstrates that mechanical wrinkling is a viable, non-invasive strategy for tuning electromechanical properties in Janus TMDC monolayers. In contrast to conventional TMDCs in their most common 2H stacking, which require explicit mechanical manipulation (e.g., strain gradients) to break their intrinsic horizontal mirror symmetry~\cite{sung2020,du2021,liang2022,wang2022l}, Janus architectures possess an intrinsic out-of-plane dipole moment that directly interferes either constructively or destructively, depending on sublayer stacking, with curvature-induced polarization.

Our geometric model connects the global polarization with the out-of-plane projection of the local sheet normal and the areal density of the formula units. In a Janus monolayer, wrinkling reorients the intrinsic dipole moment, while also modifying the local environment. The opposing chalcogen projected-weight trends and varying atomic dipoles provide a complementary microscopic description of the changes accompanying this geometric response.

By evaluating a comprehensive range of supercell sizes and strain levels, we established clear operational boundaries for continuum polarization modeling in corrugated dipolar 2D sheets. In small supercells, discrete lattice fluctuations and localized bond reorientations perturb the cancellation between metal and chalcogen dipole moments, introducing minor scattering around the master curve. Under extreme compressive strains, localized non-linear atomic relaxations take over. These high-curvature regimes mark the breakdown of linear behaviour, defining an upper physical boundary for continuum elastica approximations. Finally, smooth nanowrinkles of moderate-to-large lateral extent occupy an ideal continuum regime where macroscopic polarization scales predictably via $P_z^{\mathrm{folded}} \propto b X$.

From a material engineering perspective, these operational boundaries offer concrete design rules: by controlling substrate strain and wrinkle aspect ratio ($h/L_{\mathrm{proj}}$), one can deterministically engineer local dipoles and electronic band alignments for low-power flexible sensors, self-powered wearable energy harvesters, and mechanically tunable optoelectronic devices.

\section{Summary and Conclusions}
\label{sec:conclusion}
In summary, we combined DFT calculations with a continuum elastica model to quantify how curvature enhances polarization in Janus \ce{MoSSe}/\ce{MoSeS} nanowrinkles. By adopting a direct-density sign-folding convention, we demonstrated that the out-of-plane polarization across diverse supercell geometries is described by a linear master curve governed by the projected aspect ratio $h/L_{\mathrm{proj}}$. A geometric projection model describes the global polarization through reorientation of the intrinsic dipole moment of Janus monolayers, using local-normal projection and formula-unit areal density, with high fidelity. Atom-resolved dipoles and opposing projected spectral weight trends of S $3p$/Se $4p$ orbitals characterize the accompanying microscopic change. These results provide a basis for mechanically tunable polarization functionality in Janus monolayers.

Beyond advancing the fundamental understanding of curvature-driven electromechanical coupling in low-dimensional materials, these results establish a predictive framework for designing adaptive 2D architectures. By demonstrating that macroscopic polarization can be tuned through deterministic strain engineering and wrinkle geometry, this work opens new avenues for non-invasive domain engineering in non-centrosymmetric monolayers. The geometry-dependent polarization reported here suggests specific routes toward electronic and optoelectronic functionality. Wrinkle-induced dipole modulation in Janus structures could modify electrostatic alignment and carrier-injection barriers, influence photocarrier recombination, or modify interlayer charge separation in vertical, multilayer, aligned Janus structures. Thus, looking forward, these microscopic design rules provide a blueprint for experimental realizations of flexible nanoelectronics, atomically modulated piezotronic sensors, and strain-engineered optoelectronic devices where local dipole landscapes dictate functional electronic performance.

\section*{Acknowledgements}
Computational resources were provided by the high-performance computing system Emmy at Göttingen through the project FUSE-TOP (thp00002), supported by the German Federal Ministry of Education and Research (BMBF) and the National High-Performance Computing (NHR) program.

\section*{Data and Code Availability}
Input and output files of the ab initio calculations performed in this work are available free of charge in Zenodo DOI: 10.5281/zenodo.22250570. The code \texttt{Quantum ESPRESSO} (\href{http://www.quantum-espresso.org}{http://www.quantum-espresso.org}) used in this work is freely available. The Python packages developed in-house for post-processing and visualization are available from the corresponding author upon reasonable request.

\section*{Author Contributions}
\textbf{Stefan Velja}: conceptualization, investigation and formal analysis, writing original draft; \textbf{Surender Kumar:} investigation and formal analysis, review \& editing, writing original draft; \textbf{Domenico Corona:} investigation and formal analysis, writing – review \& editing; \textbf{Caterina Cocchi:} conceptualization, resources, supervision, writing – review \& editing.

\section*{Conflicts of Interest}
There are no conflicts of interest to declare.

%\bibliography{rsc} %You need to replace "rsc" on this line with the name of your .bib file
%\bibliographystyle{rsc} %the RSC's .bst file
\providecommand*{\mcitethebibliography}{\thebibliography}
\csname @ifundefined\endcsname{endmcitethebibliography}
{\let\endmcitethebibliography\endthebibliography}{}

\end{document}